\documentclass[letterpaper,12pt]{article}
\pdfoutput=1 

\usepackage{jheppub} 
\usepackage{amsmath,amsfonts,amssymb}
\usepackage{psfrag}
\usepackage{enumerate}
\usepackage{mathrsfs}
\usepackage{graphicx}
\usepackage{wrapfig}
\usepackage{xcolor}
\usepackage{caption}
\usepackage{amsthm}

\numberwithin{equation}{section}

\makeatletter
\def\@fpheader{\relax}
\makeatother

\newcommand{\be}{\begin{equation}}
\newcommand{\ee}{\end{equation}}
\newcommand{\beq}{\begin{eqnarray}}
\newcommand{\eeq}{\end{eqnarray}}

\def\[{\left [}
\def\]{\right ]}
\def\({\left (}
\def\){\right )}

\def\r2{\sqrt{2}}

\newcommand{\bbibitem}[1]{\bibitem{#1}\marginpar{#1}}

\newcommand{\myeq}[1]{\begin{equation} #1 \end{equation}}
\newcommand{\myal}[1]{\begin{align} #1 \end{align}}
\newcommand{\OO}{\mathcal{O}}
\newcommand{\Hm}{H_\text{mod}}

\def\Label#1{\label{#1}%
  \smash{\hbox to0pt{\raise1ex\hbox{\tiny[#1]}\hss}}}
\def\noLabels{\let\Label=\label}
\def\nobbibitem{\let\bbibitem=\bibitem}

\title{Holographic Order from Modular Chaos}

\author[a]{Jan de Boer}

\author[b]{Lampros Lamprou}
\affiliation[a]{Institute for Theoretical Physics and Delta Institute for Theoretical Physics, University of Amsterdam, PO Box 94485, 1090GL, Amsterdam, The Netherlands}
\emailAdd{j.deboer@uva.nl}
\affiliation[b]{Center for Theoretical Physics, Massachusetts Institute of Technology, Cambridge, MA 02139-4307, USA}
\emailAdd{llamprou@mit.edu}

\abstract{We argue for an exponential bound characterizing the chaotic properties of modular Hamiltonian flow of QFT subsystems. In holographic theories, maximal modular chaos is reflected in the local Poincare symmetry about a Ryu-Takayanagi surface. Generators of null deformations of the bulk extremal surface map to \emph{modular scrambling modes} ---positive CFT operators saturating the bound--- and their algebra probes the bulk Riemann curvature, clarifying the modular Berry curvature proposal of arXiv:1903.04493.}

\begin{document}
\noLabels 
\nobbibitem 

\maketitle
\flushbottom
\tableofcontents

\section{Entanglement, Chaos and Emergent Spacetime}
Gravity in asymptotically AdS Universes appears to be an effective description of the \emph{storage} and \emph{processing} of quantum information in special strongly coupled, chaotic CFTs. Geometric features of the holographic spacetime reflect the character of entanglement in this dual quantum theory \cite{Ryu:2006bv, VanRaamsdonk:2010pw, Maldacena:2013xja}, while probes of quantum chaos \cite{Maldacena:2015waa}---most famously out-of-time-ordered correlation functions--- compute classical gravitational scattering processes near black holes, with the maximal Lyapunov exponent controlled by the bulk gravitational redshift \cite{Shenker:2013pqa}. In this paper, we unify these two approaches in a notion of chaos associated to the entanglement of a state which we suggest may be the physical principle underlying the emergence of the dual spacetime's local structure: The local Poincare algebra and curvature about Ryu-Takayanagi surfaces.

Every subalgebra of quantum observables is endowed with a natural internal clock, stemming from its correlations with the rest of the degrees of freedom. This ``time'' evolution is generated by the modular Hamiltonian of the subsystem, heuristically defined as $\Hm=-\log \rho$, and it is distinct from the usual Hamiltonian flow. Inspired by the observations of \cite{Faulkner:2018faa}, we develop and prove, under certain assumptions, a bound on modular Hamiltonian chaos for Quantum Field Theories. 

Our main claim in Section \ref{sec:partI} is that any infinitesimal perturbation of the modular Hamiltonian of a QFT subregion, resulting from either a change of the global state or a shape deformation, cannot contain matrix elements that grow faster than exponentially in modular time $s$ with exponent $2\pi$, in the limit $s\to \pm \infty$. For large $N$ theories, like holographic CFTs, we propose that the same exponential bound also controls the growth of ``code subspace'' matrix elements of $\delta \Hm(s)$, at intermediate modular timescales $1\ll 2\pi s\ll \log N$
\myal{ &\lim_{\substack{1\ll N \\ 1\ll 2\pi |s|\ll \log N}} \Big| \frac{d}{ds} \log  F_{ij}(s) \Big| \leq 2\pi \nonumber\\
\text{where: } &F_{ij}(s)=\Big|\langle \chi_i| e^{i\Hm s} \delta \Hm e^{-i\Hm s} |\chi_j\rangle\Big|,  \,\,\,\, \forall\, |\chi_i\rangle \in \mathcal{H}_{\text{code}}^{\psi}.}

In Section \ref{sec:partII}, we turn this bound into a new organizing principle: We look for those operators that saturate modular chaos, which we dub \emph{modular scrambling modes}. Scrambling modes generate approximate symmetries of the reference quantum state. In holographic CFTs, by virtue of the JLMS relation \cite{Jafferis:2015del} they map to local null shifts of the bulk Ryu-Takayanagi surface $\zeta^\pm$ that preserve its normal frame and are, as a result, approximate isometries of the metric at the surface
\myeq{\mathcal{L}_{\zeta_{\pm}} g_{\mu\nu}\big|_{RT}\approx 0}
whereas they approximately commute with ``simple'' local bulk operators outside a small neighborhood of the extremal surface. These operators are the modular parallel transport generators of \cite{Czech:2019vih}. We explicitly show that their commutator probes the bulk Riemann curvature at the surface (\ref{Gcommutator}), a fact that underlies the modular Berry holonomies of \cite{Czech:2019vih}. In Section \ref{sec:conclusion}, we discuss possible connections of our construction to other recent attempts to obtain bulk translations from properties of quantum chaos and take an ambitious look at some exciting future directions.

\section{Part I: A Chaos Bound for Modular Flow}
\label{sec:partI}

In this section, we characterize the chaotic properties of the unitary flow generated by modular Hamiltonians of subregions, in the form of a precise bound, motivated by the observations of \cite{Faulkner:2018faa}. Motivated by this general property of modular flow, we argue in Section \ref{subsec:holographicbound} for a related bound for the large $N$ theories. As we explain in Section \ref{sec:partII}, this notion of modular chaos plays a central role in the emergence of the holographic spacetime.

\subsection{Background}
\label{subsec:background}
For completeness and clarity we begin with a brief review of modular Hamiltonians and some of their essential properties. Further details and proofs of the claims below can be found in \cite{witten, araki1}. 

\paragraph{The Modular Hamiltonian.} Given a quantum state $|\Psi\rangle$ in some Hilbert space $\mathcal{H}$ and a subalgebra of observables $\mathcal{A}$, the modular Hamiltonian is a Hermitian operator that encodes the entanglement pattern between the degrees of freedom in $\mathcal{A}$ and the rest of the system. In theories with spatial organization of the degrees of freedom, e.g. QFTs or lattice systems, a natural set of subalgebras of interest are constructed by operators localized within the domain of dependence of different spatial regions. 

When the Hilbert space admits a tensor factorization $\mathcal{H}=\mathcal{H}_R\otimes \mathcal{H}_{R^c}$, the complementary regions $R$, $R^c$ come equipped with reduced density matrices $\rho_R$ and $\rho_{R^c}$ respectively and the (full) modular Hamiltonian is defined as:
\myal{ \Hm &= -\log \rho_R +\log \rho_{R^c} \\
\rho_R &= \text{Tr}_{R^c}\Big[ |\Psi\rangle \langle \Psi |\Big]\nonumber\\
\rho_{R^c} &= \text{Tr}_R\Big[ |\Psi\rangle \langle \Psi |\Big]}

In relativistic Quantum Field Theories, such a tensor decomposition of the Hilbert space is not possible, as a consequence of the Reeh-Schlieder theorem when $|\Psi\rangle $ is cyclic and separating, which we assume is the case throughout \cite{witten}. Nevertheless, the modular Hamiltonian associated to the algebra $\mathcal{A}(R)$ of operators in $R$ is still a well defined, unbounded operator that annihilates the state
\myeq{ \Hm |\Psi\rangle =0\label{Hmodonpsi}}
and generates a unitary automorphism of $\mathcal{A}(R)$:
\myeq{e^{i\Hm s} \mathcal{A}(R) e^{-i\Hm s} = \mathcal{A}(R).}
Modular flow is, therefore, a natural notion of an internal clock for a subalgebra of observables. This modular time evolution is a priori unrelated to the external dynamical clock defined by the Hamiltonian of the theory.

The formal QFT definition of $\Hm$ follows the introduction of the modular conjugation $S_{\Psi}$, an anti-linear operator associated to the state $|\Psi\rangle$ and the subalgebra $\mathcal{A}(R)$, acting as:
\myal{S_\Psi O|\Psi\rangle &= O^\dagger |\Psi \rangle, \,\,\, \forall O\in \mathcal{A}(R). \label{modularconj}}
The positive operator
\myeq{\Delta_\psi =S^\dagger_\psi S_\psi \label{modularop}}
is called \emph{modular operator}. The modular Hamiltonian is, then, defined as:
\myeq{\Hm = -\log \Delta_\psi \label{hmoddef}}
A familiar example is the modular Hamiltonian for a half-space region in the vacuum state of a QFT in Minkowski background, which is simply the generator of Rindler boosts \cite{witten}.

\paragraph{Some important properties.} The modular Hamiltonian is an unbounded operator and as a result it is not defined on the entire Hilbert space. Nevertheless, its domain $\mathcal{D}\subset \mathcal{H}$ is a dense subspace of $\mathcal{H}$ and, in particular, the action of modular flow on the states $O|\Psi\rangle$, $O\in \mathcal{A}$ yields normalized Hilbert space vectors
\myal{  |O,s\rangle_\Psi= \Delta^{is}_\psi O|\Psi\rangle \,\,\text{with:}\,\, \,_\Psi\langle O,s |O,s\rangle_\Psi < \infty \label{holomorphy} }
for as long as the modular parameter $s$ lies within the complex strip $ -\frac{1}{2}\leq\text{Im}[s]\leq0$. The flowed states $|O,s\rangle_\Psi$ are therefore continuous and holomorphic in the interior of the strip. 

The finite width of the analyticity region is a general feature of modular flow in unitary, Lorentz invariant quantum field theories and will be crucial for our discussions in this paper. Analyticity in the strip's interior can be argued as follows. For $\text{Im}[s]=0$, $\Delta^{is}$ is a unitary so (\ref{holomorphy}) is trivial. Similarly, for $\text{Im}[s]=-\frac{1}{2}$ the norm of the state is:
\myeq{ \,_\Psi\langle O,s=-i/2 |O,s=-i/2\rangle_\Psi= \,_\Psi\langle O |\Delta_\psi |O\rangle_\Psi=\,_\Psi\langle O^\dagger  |O^\dagger \rangle_\Psi <\infty}
where we used definitions (\ref{modularop}), (\ref{modularconj}). The boundedness for  $-\frac{1}{2}<\text{Im}[s]<0$ follows from the general property $\Delta_\psi^r <\Delta_\psi +1$ for $0<r<1$ which holds for any positive semi-definite operator.

Consider, now, two spacetime subregions $R_1$ and $R_2$ and the associated $\Hm$'s. When the two regions are nested $R_1\subset R_2$ one can prove 3 statements we will make crucial use of in Section \ref{subsec:chaosbound}:
\begin{enumerate}
\item $\Hm(R_1) \leq \Hm (R_2)$
\item $ \Delta^{a}_\psi (R_1)\geq \Delta^{a}_\psi(R_2)$, for $0\leq a\leq 1$
\item $||  \Delta^{is}_\psi(R_2)  \Delta^{-is}_\psi(R_1) || \leq 1$     for     $-\frac{1}{2}\leq \text{Im}[s]\leq 0, \,\text{Re}[s]\in \mathbb{R}$.
\end{enumerate}
Conditions 1 and 2 are operator inequalities and they underly the algebraic proof of monotonicity of relative entropy \cite{araki1, witten}. The last condition can straightforwardly be derived from condition 2 and holomorphy (\ref{holomorphy}) and it implies that the operator $ \Delta^{is}_\psi(R_2)  \Delta^{-is}_\psi(R_1)$ is holomorphic in the interior of the complex strip.

\subsection{The modular chaos bound}
\label{subsec:chaosbound}
Equipped with the above concepts, we are ready to state the bound on modular chaos. Consider a state $|\Psi\rangle$ and the modular Hamiltonian associated to the subalgebra $\mathcal{A}(R)$ of subregion $R$. We can infinitesimally perturb the modular Hamiltonian $\Hm \rightarrow \Hm +\epsilon \delta\Hm$ in two ways: by a deformation of the region's shape $R+\epsilon\delta R$ or by a small change in the global state $|\Psi\rangle \rightarrow |\Psi\rangle +\epsilon \delta |\Psi\rangle$. The modular chaos bound states that, under the original modular flow, the matrix elements of $\delta \Hm$ grow at most exponentially in $s$, with an exponent bounded by $2\pi$:

\myal{ &\lim_{s\to \pm \infty} \Big|\frac{d}{ds} \log  F_{ij}(s)\Big| \leq 2\pi \nonumber\\
\text{where: } &F_{ij}(s)=\Big|\langle \chi_i| e^{i\Hm s} \delta \Hm e^{-i\Hm s} |\chi_j\rangle\Big|,  \,\,\,\, \forall\, |\chi_i\rangle \in \mathcal{D} \label{thebound}}
In subsections \ref{subsubsec:shape}-\ref{subsubsec:state} we present evidence in support of this bound. We then use holographic intuition to conjecture a stronger variant of (\ref{thebound}) for large $N$ theories ---a claim we will utilize for our holographic application in Section \ref{sec:partII}.

\subsubsection{Shape deformations}
\label{subsubsec:shape}

\paragraph{Nested deformations.} Our first argument in favor of (\ref{thebound}) comes from considering a shape deformation where $R+\epsilon \,\delta R\subset R$. As reviewed in Section \ref{subsec:background}, this nesting property guarantees that the operator $e^{-i\Hm s}e^{i(\Hm+\epsilon \delta \Hm) s}$ is bounded and holomorphic in $s$:
\myeq{\Big|\Big| e^{-i\Hm s}e^{i(\Hm+\epsilon \delta \Hm) s}\Big|\Big| \leq 1 \label{boundedness}}
in the $-\frac{1}{2}\leq \text{Im}[s]\leq 0$ strip. As we now explain, the modular chaos bound follows from the finite width of this analyticity strip, which is itself a consequence of unitarity and Lorentz invariance. 

The two exponentials in (\ref{boundedness}) can be combined using the Baker-Campbell-Haussdorff identity to obtain, at leading order in $\epsilon$:
\myal{ e^{-i\Hm s}e^{i(\Hm+\epsilon \delta \Hm)s} &= \exp \Big[ i\epsilon \left(s \,\delta \Hm -\frac{i s^2}{2}[\Hm, \delta \Hm] +\frac{i^2 s^3}{3!} [\Hm, [\Hm, \delta \Hm]] +\dots \right) \nonumber\\
&+\OO(\epsilon^2)\Big]\nonumber\\
&= \exp \Big[ i\epsilon \int_0^s ds'\, e^{-i\Hm s'}\delta \Hm e^{i\Hm s'} +\OO(\epsilon^2)\Big]\label{BCH}}
To make contact with the bound (\ref{thebound}) we must introduce further assumptions about the matrix elements of $\delta \Hm(s)$ as functions of complex modular time $s$. The simplest derivation of (\ref{thebound}) is achieved by assuming that for $\text{Re}[s]\gg 1$ the dominant contribution to $\langle \chi_i|\delta \Hm(s) |\chi_j\rangle$ is an exponential of $s$
\myeq{e^{-i \Hm s} \delta \Hm e^{i\Hm s} \sim e^{\lambda s} G_+ \label{growth}}
where $\lambda>0$ and $G_+$ is a Hermitian operator encoding the relevant ``stripped''\footnote{without the exponential prefactor.} matrix elements of $\delta \Hm (s)$ at large $s$, for which consistency requires $[\Hm, G_+]\approx i\lambda G_+$. This is a strong assumption because it constrains the functional dependence of $\delta \Hm(s)$ on the imaginary part of $s$ at late real modular times, but we will relax it below. 

In view of (\ref{growth}) and (\ref{BCH}), the boundedness condition (\ref{boundedness}) at $s=-\frac{i}{2} +\sigma$ with $\sigma \gg 1$ translates to 
\myal{ \Big|\Big|\exp \Big[ \frac{i\epsilon\, e^{\lambda \sigma}}{\lambda}(\cos\frac{\lambda}{2} -i\sin \frac{\lambda}{2} ) G_+ \Big] \Big|\Big| &\leq 1 \nonumber\\
\Rightarrow \, \Big|\Big|\exp \Big( \frac{\epsilon\, e^{\lambda \sigma}}{\lambda}\sin \frac{\lambda}{2} \,G_+ \Big) \Big|\Big| &\leq 1 \label{boundlaststep}}
where in the second line we exploited the Hermiticity of $G_+$. Our conjectured bound on the growth of $\delta \Hm$ under modular flow follows directly from (\ref{boundlaststep}), once we recall that for nested spacetime regions $R+\epsilon \delta R \subset R$ we have $\delta \Hm \leq 0$ which by virtue of (\ref{growth}) implies $G_+\leq 0$. For condition (\ref{boundlaststep}) to be satisfied we, therefore, need:
\myeq{ \lambda \leq 2\pi \label{maxlyapunov}}
up to integer multiples of $4\pi$. The same argument for the boundedness of (\ref{boundlaststep}) for other imaginary modular time values within the strip, then, excludes the possibility of integer multiples of $4\pi$.

Equations (\ref{growth}), (\ref{maxlyapunov}) constitute the bound (\ref{thebound}) on modular chaos and illustrate that the magnitude of the maximal ``Lyapunov'' exponent is controlled by the width of the complex strip where (\ref{boundedness}) holds. The same argument can be applied for $\text{Re}[s] \rightarrow -\infty$, in which case the dominant contributions to $e^{-i\Hm s} \delta \Hm e^{i\Hm s}$ come from operators $[\Hm, G_-]\approx-i \lambda G_-$ where $\lambda$ must again be upper bounded by $2\pi$.

\paragraph{Relaxing assumption (\ref{growth})} In the proof above, we had to assume exponential behavior on the
entire strip, not just near the real axis. We can also make an argument
assuming only exponential behavior near the real axis as follows. Recall
that we are studying functions $f(s)$ that are analytic and bounded on the
strip $-\frac{1}{2}\leq \text{Im}[s]\leq 0$. Using $z=e^{2\pi s}$, we
can map this strip to the lower half plane, so that we might as well
study bounded functions on the lower half plane instead. We now want to
use the Kramers-Kronig relations to express the function in terms of an
integral along the real axis. To do this, the function needs to be
analytically extendible slightly away from the strip. This is in general
not possible, but this issue can be circumvented by considering
$z=i\epsilon + e^{2\pi s}$ so that the strip maps to the upper half plane
plus a tiny strip of width $\epsilon$. Second, the Kramers-Kronig
relations can only be applied if the relevant functions decays
sufficiently fast at
infinity. While it seems reasonable to assume that it does, as we expect
the relevant matrix elements to decay at late modular times, we do
not have a proof of this statement. Even if some of the matrix elements
were to approximate a constant at infinity, we can treat the constant
separately, and this would not interfere with the rest of this analysis.
One can then show with a straightforward application of
the Cauchy theorem and the Kramers-Kronig relations that
\be \label{kkrelation}
f(z) = -\frac{1}{\pi} \int_{-\infty}^{\infty} dz'
\frac{\text{Im}\,f(z')}{z'-z}
\ee
where values on the real axis are obtained by taking the limit from the
lower half of the complex plane. We will be interested in $f(s)$ that
arise by considering diagonal matrix elements: $\langle \chi |
e^{-i\Hm(R_2)s}e^{i\Hm(R_1)s}| \chi \rangle=1+if(s)$. Clearly, $f(s)$ is
bounded and $\text{Im}\, f(s)\geq 0$ because the norm of the operator is
bounded by one. If we take equation (\ref{kkrelation}) and subtract
its complex conjugate we get an equation for just $\text{Im}\, f(s)$.
Differentiating that equation as in \cite{caron} then leads to the inequality
\be
y\partial_y \text{Im}\, f(z) \leq \text{Im}\, f(z)
\ee
where $z=x+iy$.

For infinitesimal deformations, these equations remain valid, except
that we now write the matrix element as $1+i\epsilon f(s)$ so that $f(s)$
no longer need to be bounded. Converting the above equation back into
the original variables $s$ we obtain
\be
\left(
\frac{e^{2\pi s}-e^{2\pi \bar{s}}}{2 i} \frac{i}{2\pi} (e^{-2\pi
s}\partial_s - e^{-2\pi\bar{s}} \partial_{\bar{s}}) \right)
{\rm Im}\, f(s) \leq {\rm Im}\, f(s)
\ee
If we then insert $f(s) \sim g_0 e^{\alpha s}$ with real $g_0,\alpha$
into this, and take $s=r+it$, then
\be
g_0 \frac{\alpha e^{\alpha r}}{2\pi} \cos((\alpha-2\pi)t) \sin 2\pi t
\leq  g_0 e^{\alpha r} \sin \alpha t
\ee
where the right hand side needs to be non-negative as well. This latter
fact requires that $g_0$ is negative (which follows from
the fact that $\delta H$ is a negative operators) and also require that
$\alpha \leq 2\pi$. With $g_0$ negative, the remaining
inequality can also be written as
\be
  \alpha\cos((\alpha-2\pi)t) \sin 2\pi t \geq  2\pi \sin \alpha t
\ee
One can check that this inequality holds for $\alpha \leq 2\pi$ but is
violated for $\alpha > 2\pi$. The violation now already
happens for $t$ near zero, and we do not have to go all the way to
$t=-1/2$. Indeed, expanding the left and right hand side
to third order in $t$ we get
\be
2 \alpha \pi t + (-\alpha^3 \pi + 4 \alpha^2 \pi^2 -\frac{16}{3} \alpha
\pi^3)t^3 + \ldots \geq 2 \alpha\pi t
- \frac{1}{3} \alpha^3 \pi t^3 + \ldots
\ee
and the difference is
\be
\frac{2}{3} \alpha \pi (2\pi-\alpha)(4\pi-\alpha) (-t)^3 + \ldots
\ee
and which indeed changes precisely at $\alpha=2\pi$.

 \paragraph{General shape deformations} A useful property of (\ref{thebound}) is that it bounds the first order variation of the modular Hamiltonian which is linear in the deformation parameter. As a result, while the boundedness property (\ref{boundedness}) ---which played a central role in the proof of (\ref{thebound}) above--- only holds for nested spacetime regions $R+\epsilon \delta R\subset R$, the bound (\ref{thebound}) is automatically satisfied for all shape deformations of $R$. 

The reason is simple: The diffeomorphism generating an arbitrary infinitesimal shape deformation $\delta R$ can always be expressed as the linear combination of two contributions $\delta R_1$ and $\delta R_2$ for which $R+\epsilon \delta R_1 \subset R$ and $(R +\epsilon \delta R_2)^c \subset R^c$. A straightforward application of the above proof for $\delta R_1$ and $\delta R_2$ separately will bound the growth of the corresponding variations of the modular Hamiltonian, $\delta H_1$ and $\delta H_2$. Then it directly follows that (\ref{thebound}) will also bound the sum of the two.

\paragraph{Caveat} The above argument does not exclude possible faster-than-exponential growth of the modular matrix elements. It instead bounds the exponent after assuming at-most exponential growth. It will be interesting to look for a stronger argument that forbids faster growth, or find counter-examples.

\subsubsection{Virasoro excitations in CFT$_2$}
\label{subsubsec:virasoro}
An example where we have analytic control over subregion modular Hamiltonians is the CFT$_2$ vacuum and its Virasoro excitations. The modular Hamiltonian of a connected interval in the vacuum of the CFT on $S^1$ is an element of the global conformal algebra which keeps the endpoints of the interval $(x^+_L, x^-_L)$, $(x^+_R, x^-_R)$ fixed and boosts the timeslice. Without loss of generality we can take this interval to be half of the circle and its modular Hamiltonian reads:
\myal{\Hm &=K+\bar{K}\nonumber\\
K&=\pi i (L_1-L_{-1})= -2\pi i \sin x^+ \partial_+ \label{cft2K}\\
\bar{K}&=\pi i (\bar{L}_1-\bar{L}_{-1})= -2\pi i \sin x^- \partial_-}
The decomposition of $\Hm$ into a left and right moving piece comes from the holomorphic factorization of the conformal group and its normalization is fixed by requiring that, near the endpoints, it becomes $2\pi$ times the boost generator

An infinitesimal Virasoro excitation will perturb the modular Hamiltonian by some element of the Virasoro algebra $Y+\bar{Y}$ that can be expressed as a stress tensor integral against a function $f(x^+)$ that is regular everywhere on the circle:
\myeq{Y=\int dx^+ f(x_+) T_{++}(x^+)}
with a similar expression for $\bar{Y}$. Using (\ref{cft2K}) and the known Virasoro algebra, we can organize Virasoro generators in eigenoperators of $K$
\myeq{ [K, Y_{\lambda}]=2\pi i \lambda Y_{\lambda} \label{cft2eigenop}}
and use them as an operator basis for expressing the general operator $Y$ above. A straightforward calculation leads to the following expression for $Y_\lambda$:
\myeq{Y_\lambda= \int dx^+(1-\cos x^+)^{\frac{1+\lambda}{2}} (1+\cos x^+)^{\frac{1-\lambda}{2}} T_{++}(x^+).\label{ylambda}}
It is obvious from (\ref{ylambda}) that only eigenoperators with $\lambda=0, \pm 1$ correspond to smooth, $C^\infty$ diffeomorphisms of the circle.
If there would be an $Y$ such that it would grow as $e^{\alpha|s|}$ under modular flow as $s\rightarrow \pm \infty$, then the corresponding $f$
needs to behave as $(1-\cos x^+)^{\frac{1+\alpha}{2}} (1+\cos x^+)^{\frac{1-\alpha}{2}}$ near $x^+=0$ or $x^+=\pi$, as modular flow 
moves points to the endpoints of the interval, and if $\alpha$ would be larger than $2\pi$ the function $f$ must necessarily 
have singularities at either $x^+=0$ or $x^+=\pi$. Singular $f$ do not correspond to bona fide Virasoro generators and this shows that
$\alpha \leq 2\pi$. 

It then follows from (\ref{cft2eigenop}) that the bound (\ref{thebound}) is not only satisfied but it can also be saturated. 
The operators that saturate the bound are, in fact, elements of the global conformal algebra. This transformation does not change the CFT 
state but it moves the subregion that supplies $\Hm$, thus we learn that the modular chaos bound can be saturated for shape deformations.
The explicit expression of the operators that saturate the bound is $Y_{\pm 1}=L_1+L_{-1}\pm 2 L_0$. By the general results of the previous
section these must be positive operators and one can indeed explicitly check that the expectation value of $Y_{\pm 1}$ is non-negative in
any CFT state. If we map a diamond on the cylinder to the plane, $Y_{\pm 1}$ can be see to be equal to ANEC operators $\int dx^+ T_{++}$, which
is an example of a more general relation  between operators that saturate the 
modular chaos bound, ANEC operators and shocks. We will revisit this connection in the concluding section.

\paragraph{Violation of the bound for higher spin theories}
It is straightforward to repeat the analysis in the presence of a higher spin field.
For a field with spin $h$ we get
\myeq{Y_\lambda= \int dx^+(1-\cos x^+)^{\frac{h-1+\lambda}{2}} (1+\cos x^+)^{\frac{h-1-\lambda}{2}} W_h(x^+)}
which leads to a bound $\lambda \leq 2\pi(h-1)$. Interestingly, this is in perfect 
agreement with the violation of the chaos bound in higher spin theories
found in \cite{perl}. It would be interesting to analyze this further, in particular it is not clear which types of higher spin
deformations can actually arise as changes of modular Hamiltonians.

\subsubsection{State perturbations}
\label{subsubsec:state}

We will now present evidence for the modular chaos bound for infinitesimal state perturbations obtained by acting on $|\psi\rangle$ with bounded operators $A\in \mathcal{A}$. The derivation is analogous to the shape deformation discussion above. The key new ingredient is the boundedness of the analytic continuation of the \emph{unitary cocycle} for infinitesimal perturbations (\ref{cocyclebound}), a notion we now introduce in detail.

\paragraph{The unitary cocycle} A remarkable and rather counter-intuitive result of algebraic QFT is that, for a given subalgebra $\mathcal{A}$, the modular flows associated to any two global states $|\psi_1\rangle,\, |\psi_2\rangle \in\mathcal{H}$\footnote{The global states could also be mixed.} are related by an inner automorphism. Namely, there exist unitary operators $U_{\psi_2\psi_1}(s)\in \mathcal{A}$ that depend continuously on $s$ with the property 
\myeq{U_{\psi_2\psi_1}(s)\Delta^{is}_{\psi_1} O \Delta^{-is}_{\psi_1}U_{\psi_2\psi_1}^\dagger(s)=  \Delta^{is}_{\psi_2} O \Delta^{-is}_{\psi_2} \,\,\,\,  \forall O\in \mathcal{A} \label{cocycle}}
The operators $U_{\psi_2\psi_1}(s)$ define the \emph{unitary cocycle} \cite{araki1}. A similar cocycle $U_{\psi_2\psi_1}'(s)\in\mathcal{A}'$ can be defined for the modular flow of the commutant algebra $\mathcal{A}'$. The properties of the cocycle will play a key role in the chaos bound derivation below.

To understand the cocycle better we need to introduce the relative modular operator \cite{witten}. Given the states $|\psi_1\rangle,\, |\psi_2\rangle$ we can define the relative modular conjugation $S_{\psi_2\psi_1}$ as the anti-linear operator satisfying:
\myeq{S_{\psi_2\psi_1}O|\psi_1\rangle= O^\dagger |\psi_2\rangle. \label{relativeconjugation}}
In close analogy to the modular operator (\ref{modularop}), the relative modular operator is then defined as
\myeq{\Delta_{\psi_2\psi_1}=S^\dagger_{\psi_2\psi_1}S_{\psi_2\psi_1}. \label{relativemodularH}}
It is a positive semi-definite operator with $\Delta_{21}^z$ holomorphic and bounded for $-\frac{1}{2}\leq \text{Im}[z]\leq 0$ and with the properties:
\myal{&\Delta_{\psi_2\psi_1}^{1/2}|\psi_1\rangle= |\psi_2\rangle. \label{relativemodularonpsi}\\
&\Delta_{\psi_2\psi_1}=\Delta_{\psi_1} \,\,\text{when: } |\psi_1\rangle= |\psi_2\rangle\label{relative1eq2}\\
&S(\psi_2|\psi_1) =-\langle \psi_1| \log \Delta_{\psi_2\psi_1}|\psi_1\rangle >0 \label{relativeentropy}}
where $S(\psi_2|\psi_1)$ is the relative entropy of the corresponding reduced states on the subalgebra $\mathcal{A}$.

The unitary cocycle can then formally be constructed in terms of the modular and relative modular flows
\myeq{U_{\psi_2\psi_1}(s)= \Delta_{\psi_2\psi_1}^{is}\Delta^{-is}_{\psi_1}. \label{cocycledef}}

The combination of eq. (\ref{cocycle}) and (\ref{cocycledef}) reveals that the relative modular flow acts as the modular flow of $\Hm^2$ on operators in $\mathcal{A}$. On the other hand, because $U_{\psi_2\psi_1}\in \mathcal{A}$ ---and therefore commutes with operators in the commutant $\mathcal{A}'$---  $\Delta^{is}_{\psi_2\psi_1}$ acts as the modular flow of $\Hm^1$ on operators in $\mathcal{A}'$. The converse is true for the relative modular flow of the commutant algebra $\mathcal{A}'$.

\paragraph{Boundedness of the cocycle in the strip} The first step in arguing for the modular chaos bound for state excitations is to establish the boundedness of the unitary cocycle when analytically continued in the interior of the strip $-1/2\leq \text{Im}[s]\leq 0$:
\myeq{\big|\big| U_{\psi_2\psi_1}(s)\big|\big| \leq \max \{\kappa,1\}\label{cocyclebound}}
when the two states are related by $|\psi_2\rangle = A|\psi_1\rangle $ where $A\in \mathcal{A}$ is a bounded operator with norm:
\myeq{\big|\big|A\big|\big|=\sup_{|\phi_1\rangle,|\phi_2\rangle\in \mathcal{H}}\big|\langle\phi_1|A|\phi_2\rangle\big|=\kappa>0}

The proof \cite{araki1} is quite straightforward. Consider the action of the unitary cocycle on the dense subspace of the Hilbert space $|O'\rangle=O'|\psi_1\rangle$ obtained by acting with $O'\in\mathcal{A}'$ on the cyclic, separating vector $|\psi_1\rangle$:
\myeq{U_{\psi_2\psi_1}(z)|O'\rangle=O' U_{\psi_2\psi_1}(z)|\psi_1\rangle \label{cocycleonstates}}
When $\text{Im}[z]=0$, $U_{21}(z)$ is unitary and (\ref{cocyclebound}) is obviously satisfied. When $z=s-\frac{i}{2}$, using (\ref{relativemodularonpsi}) and (\ref{Hmodonpsi}) in (\ref{cocycleonstates}) gives:
\myal{U_{\psi_2\psi_1}(s-i/2)|O'\rangle&= O' \Delta_{\psi_2\psi_1}^{is+1/2}\Delta^{-is-\frac{1}{2}}_{\psi_1}|\psi_1\rangle = O' \Delta_{\psi_2\psi_1}^{is} |\psi_2\rangle\nonumber\\
&=O' U_{\psi_2\psi_1}(s) \Delta^{is}_{\psi_1}A \Delta^{-is}_{\psi_1} |\psi_1\rangle\nonumber\\
&=  U_{\psi_2\psi_1}(s) \Delta^{is}_{\psi_1}A \Delta^{-is}_{\psi_1}|O'\rangle}
It immediately follows that the norm of the resulting state satisfies:
\myeq{\big|\big|U_{\psi_2\psi_1}(s-i/2) |O'\rangle\big|\big| \leq \kappa \big|\big|\, |O'\rangle\big|\big| \label{boundednessihalves}}
The boundedness of $\Delta_{\psi_2\psi_1}^z$ in the strip, by virtue of the Hadamard three-lines theorem implies:
\myeq{\big|\big|U_{\psi_2\psi_1}(z) \big|\big| \leq \kappa^{-2\text{Im}[z]}\leq \max\{\kappa, 1\}}

\paragraph{The modular chaos bound} Consider now two states $|\psi_1\rangle$ and $|\psi_2\rangle = A|\psi_1\rangle$, $A\in \mathcal{A}$ where we take $||A||\leq1+\epsilon \mu$, $\mu \sim O(1)$ and $\langle \psi_1|A|\psi_1\rangle=1-O(\epsilon^2)$, $\epsilon \ll 1$ so that the new state is an infinitesimal perturbation of $|\psi_1\rangle$. From eq. (\ref{cocycle}) we know that
\myeq{\big| \big| \Delta_{\psi_2\psi_1}^{is}\Delta^{-is}_{\psi_1}\big|\big| =\big|\big| U_{\psi_2\psi_1}(s)\big|\big|\leq 1+\epsilon \mu \label{perturbativeboundedness}}
as long as $-1/2\leq \text{Im}[s]\leq 0$. Recalling (\ref{relative1eq2}), we can perturbatively expand the relative modular flow as: $\Delta_{\psi_2\psi_1}^{is}\approx \exp[-i(\Hm^1 +\epsilon \delta D_{\psi_2\psi_1})s]$. Substituting it in (\ref{perturbativeboundedness}) we obtain an expression identical to (\ref{boundedness}), so we can repeat the procedure for shape deformations: Combine the two exponentials using BCH, assume that $\delta D_{\psi_2\psi_1}(s) \sim e^{\lambda s}G_+$ for $\text{Re}[s]\gg 1$ and take the limit $s\to \sigma -\frac{i}{2}$, $\sigma \to\infty$. As before, this translates to the requirement
\myeq{\big|\big| \exp\left[-\epsilon \frac{e^{\lambda \sigma}}{\lambda} \sin\frac{\lambda}{2} G_+\right]\big|\big| \leq \exp[\epsilon \mu]. \label{finalstep}}
If we take $\sigma \gg \frac{1}{\lambda} \log (\lambda\,\mu)$ the deviation of the right hand side from 1 becomes negligible and the boundedness of the unitary cocycle implies that:
\begin{align}
G_+&\geq 0\nonumber\\
\lambda &\leq 2\pi \label{statechaosbound}
\end{align}
which is, again, the modular chaos bound, albeit for the evolution of the relative modular operator and not the perturbation of the modular Hamiltonian. In view of (\ref{cocycle}), however, this constrains $\delta \Hm$ as well, since for every operator $O\in \mathcal{A}$ we have
\myeq{[\delta D_{\psi_2\psi_1}(s) , O] = [\delta \Hm(s), O]}
Thus, as far as the action on the elements of the subregion's algebra is concerned, the two operators are equivalent and (\ref{statechaosbound}) bounds the dominant contributions to $\delta \Hm(s\to \infty)$. It is worth emphasizing that this does not, strictly speaking, bound the matrix elements of $\delta \Hm$ as in (\ref{thebound}), but instead the matrix elements of the commutator of $\delta \Hm$ with operators $O\in \mathcal{A}$. This is actually sufficient for our holographic considerations in Section \ref{sec:partII}, since our arguments rely on this weaker modular chaos condition. Nevertheless, it would be interesting to find direct evidence for the stronger bound (\ref{thebound}) for state deformations.

It is, also, important to note that it is generally not possible to bound the action of $\delta \Hm$ on operators in the commutant $O'\in \mathcal{A}'$ using the arguments presented here. The reason is that the unitary cocycle for $\mathcal{A}'$ is bounded when $|\psi_1\rangle=A'|\psi_2\rangle$ for some $A'\in \mathcal{A'}$ with $||A'||=1+\epsilon \mu'$ which is clearly not generally true, given that we already assumed $|\psi_2\rangle=A|\psi_1\rangle$, $A\in\mathcal{A}$ with $||A||=1+\epsilon \mu$. For these special cases, however, our argument applies to the complementary region as well, constraining the growth of matrix elements of $[\delta \Hm, O]$ for any operator $O$ of the theory.

\paragraph{Exponential growth from state perturbations} A natural question is whether state perturbations can ever generate exponentially growing $\delta \Hm(s)$ in order for our modular chaos bound to have non-trivial content. To address this, let us contrast two classes of bounded operators $A$ we can use to infinitesimally perturb the state: 1) Operators $A=1+\epsilon O$ with $O$ a bounded operator $||O||=\mu$ of the subregion algebra $\mathcal{A}$, 2) operators $A=\exp[-\epsilon C]$ where $C\geq 0$ a positive semi-definite, unbounded operator. 

For the first case, the modular chaos bound is trivial: There are no possible exponential contributions to $\delta \Hm(s)$ as we send $s\to \infty$. To see this, let us first assume that indeed $\delta \Hm(s) \sim e^{\lambda s} G_+$ as before. Then (\ref{statechaosbound}) tells us that $G_+\geq 0$. However, we can also consider the state perturbation by $A'=1-\epsilon O$. The boundedness of $O$ implies the same for $A'$, so the cocycle satisfies (\ref{perturbativeboundedness}) and the chaos bound applies again. However, at first order in $\epsilon$, the change of the modular Hamiltonian is now $\delta \Hm'=-\delta \Hm$, so (\ref{finalstep}) yields $G_+\leq 0$. The cocycle boundedness for these two perturbations implies that the assumed exponentially growing matrix elements are in fact zero.

The situation is different, however, for the excitation $\exp[-\epsilon C]|\psi\rangle$ where $C$ is a positive semi-definite, \emph{unbounded} operator. The bound (\ref{statechaosbound}) applies to this state deformation but it cannot be applied to the one with the opposite sign because $\exp[\epsilon C]$ is \emph{unbounded} so the reasoning we used in our proof above is invalid. Such state perturbations can thus potentially generate exponentially growing matrix elements.

\subsection{The modular chaos bound at large $N$}
\label{subsec:holographicbound}

When the quantum field theory of interest is a large $N$ QFT, e.g. a holographic CFT, $1/N$ serves as an additional perturbative parameter. This enables us to achieve a finer characterization of modular chaos, tailored to large N systems, by considering the large $N$ limit of the general bound (\ref{thebound}). 

This limit needs to be taken with care. Instead of arbitrary matrix element of $\delta \Hm (s)$, we will consider excitations of the reference state $|\psi\rangle$ by an $n\sim O(1)$ number of low dimension operators $\Delta \sim O(1)$; that is simply the requirement that the energy difference between the states considered and $|\psi\rangle$ does not scale with $N$. In holography, when $|\psi\rangle$ is a sufficiently semi-classical state ---i.e. a state prepared by a Euclidean path integral with sources turned on--- this subspace of the Hilbert space is the, so-called, \emph{code subspace}, dual to bulk supergravity excitations about the classical spacetime generated by $|\psi\rangle$. Moreover, we want to take the large $N$ limit before we study the matrix elements of $\delta \Hm(s\to \infty)$. The exponentials $e^{\lambda s}$ that may appear in $\delta \Hm(s)$ due to modular chaos must not compete with its $1/N$ expansion, thus we must restrict $s$ to be in the intermediate modular time regime $1\ll 2\pi s\ll \log N$, namely to not scale with $N$.

We, therefore, propose that for large $N$ systems
\myal{ &\lim_{\substack{1\ll N \\ 1\ll 2\pi |s|\ll \log N}} \Big| \frac{d}{ds} \log  F_{ij}(s) \Big| \leq 2\pi \nonumber\\
&\text{where: } F_{ij}(s)=\Big|\langle \chi_i| e^{i\Hm s} \delta \Hm e^{-i\Hm s} |\chi_j\rangle\Big|,  \,\,\,\, \forall\, |\chi_i\rangle \in \mathcal{H}_{\text{code}}^{\psi} \label{theholographicbound}}
We will refer to this as the the following \emph{large $N$ modular chaos bound}.

Since (\ref{theholographicbound}) can be obtained by taking the large $N$ limit of the general bound (\ref{thebound}), the arguments presented in the previous Sections can also be used in support of the large $N$ bound: The product of modular flows for shape deformations of the region (\ref{boundedness}), or the unitary cocycle (\ref{cocycle}) for the state perturbations of section \ref{subsubsec:state}, are bounded in the entire $-1/2\leq \text{Im}[s]\leq 0$ strip. Therefore, by expanding in $N$ and assuming at most an exponential dependence of $\delta\Hm$ on $s$ for $1\ll 2\pi s\ll \log N$, we can again bound the exponent using the same arguments. 

However, the two bounds are physically distinct. Saturation of one does not imply saturation of the other; moreover, even when both are saturated, the maximally growing contributions to $\delta \Hm$ in the corresponding modular time regimes are not a priori related. It is the large $N$ modular bound (\ref{theholographicbound}) that will be important in our holographic discussion in Section \ref{sec:partII}.

In AdS/CFT, the code-subspace projection of the CFT modular Hamiltonian is mapped to the bulk modular Hamiltonian for the degrees of freedom in the entanglement wedge by the JLMS relation \cite{Jafferis:2015del, Harlow:2016vwg}
\myeq{\langle \chi_i|  \Hm^{CFT}  |\chi_j\rangle=\langle \chi_i|  \Hm^{bulk}  |\chi_j\rangle,   \,\,\,\, \forall\, |\chi_i\rangle \in \mathcal{H}_{\text{code}}^{\psi}}
Eq. (\ref{theholographicbound}) then is the boundary avatar of the modular chaos bound (\ref{thebound}) for the bulk QFT on the classical background sourced by the reference state $|\psi\rangle$. This holographic interpretation of the large $N$ bound can be viewed as an alternative motivation for (\ref{theholographicbound}) assuming one accepts the validity of the general asymptotic bound (\ref{thebound}). In the following section we will turn (\ref{theholographicbound}) into a powerful new tool for probing the local structure of the holographic spacetime.

The large $N$ modular chaos bound is a generalization to the Maldacena-Shenker-Stanford bound \cite{Maldacena:2015waa}. When the state $|\psi\rangle $ is the QFT vacuum and the subregion of interest is half of space the modular Hamiltonian is simply the Rindler boost generator. Condition \ref{theholographicbound}, therefore, bounds the large boost limit of $\delta \Hm$. By appropriately selecting our state perturbation we can generate changes in the modular Hamiltonian that are products of bounded operators in the left and right Rindler wedge $\delta \Hm =O_LO_R-\langle O_LO_R\rangle$ ---where the subtraction of the expectation value is required for recovering the normalization of the modular operator. This is, in fact, the MSS chaos bound on the Rindler OTOCs.

\section{Part II: Modular Chaos and Local Structure of the Hologram}
\label{sec:partII}

\subsection{Modular Scrambling Modes}
\label{subsec:scramblingmodes}

The bound of the previous section bans operators whose modular growth rate exceeds (\ref{theholographicbound}) from contributing to changes in $\Hm$. This renders operators that marginally comply with modular law physically interesting objects. We dub the latter \emph{modular scrambling modes}.

The modular scrambling modes cannot generally be constructed microscopically ---except in very special cases, e.g. in the CFT$_2$ vacuum (\ref{ylambda}). Nevertheless, they can be extracted from a shape variation of the modular Hamiltonian, as the dominant operator contributions in the large modular time limit:
\myal{G_+ &=\frac{1}{2\pi } \lim_{s\rightarrow \Lambda} e^{-2\pi s} e^{-i\Hm s} \delta \Hm e^{i\Hm s}\nonumber\\
G_- &=-\frac{1}{2\pi }  \lim_{s\rightarrow -\Lambda} e^{2\pi s} e^{-i\Hm s} \delta \Hm e^{i\Hm s} \label{scramblingmodes}}
where we take $1\ll 2\pi \Lambda \ll \log N$ as we explained in Section \ref{subsec:holographicbound}. $G_\pm$ obey by construction:
\myeq{ [\Hm, G_\pm] \approx \pm 2\pi i \,G_\pm \label{chaoscommutator}}
which we will refer to as the \emph{modular chaos commutator}. 

Importantly, the modular scrambling modes generate approximate symmetries of the background state $|\psi\rangle$. Invariance of $|\psi\rangle$ under modular flow of any subregion $R$ implies that $\delta \Hm|\psi\rangle=0$ which by virtue of (\ref{scramblingmodes}) becomes
\myeq{G_\pm |\psi\rangle\approx 0. \label{Gsymmetry}}
In this Section, we utilize (\ref{scramblingmodes}), supplemented by the standard holographic dictionary, to derive the dual of the modular scrambling modes and argue that the symmetry algebra (\ref{chaoscommutator}), (\ref{Gsymmetry}) is a holographic manifestation of the local Poincare symmetry of the bulk. 

More precisely, we show that, for a shape deformation 
\begin{enumerate}
\item $G_\pm$ are supported in an $O(e^{-2\pi\Lambda} \ell)$ neighborhood of the Ryu-Takayanagi surface, where  $\ell$ is the scale of non-locality of the bulk modular flow, typically set by the normal extrinsic curvature of the surface $\ell\sim 1/K_{ij|n}$, and
\item in a small neighborhood $x^\pm\ll e^{-2\pi \Lambda} \ell$ about the RT surface,\footnote{$x^\pm$ are the normal null coordinates about the RT surface located at $x^\pm=0$} $G_\pm$ are geometric transformations $\zeta^\mu_{(\pm)}$ that deform the bulk extremal surface. In particular, they are shifts along the future/past null directions that preserve the RT surface's normal frame (\ref{scramblingmodeapp}):
\myal{\mathcal{L}_{\zeta_{(\pm)}}g_{\alpha \mu}(x^\alpha=0, y^i)&=0 \label{paralleltransport}}
where $\mu$ is a bulk spacetime index, $x^\alpha$, $\alpha=0,1$ parametrize distances along the orthogonal 2D plane with the RT surface at $x^\alpha=0$, and $y^i$, $i=1,\dots, d-2$ are internal surface coordinates. Consistent with (\ref{Gsymmetry}), these are \emph{approximate local isometries} of the metric:
\myeq{\mathcal{L}_{\zeta_{(\pm)}}g_{\mu\nu}(x^\alpha=0, y^i)= O(e^{-2\pi\Lambda}) \label{localapproxsym}}
The vector field $\zeta=\zeta_{(+)}+\zeta_{(-)}$ generates the \emph{modular parallel transport} of \cite{Czech:2019vih} and is discussed further in the Appendix.
\end{enumerate}

We propose that the shift generators  (\ref{paralleltransport}) are the only bulk modular scrambling modes. For an arbitrary state or shape perturbation of the region, the large modular time limit in (\ref{scramblingmodes}) allows us to ``zoom in'' a small neighborhood of the RT surface and extract the geometric piece of $\delta \Hm$: The approximate local bulk symmetry moving the RT surface to its location.


\subsection{The bulk derivation} 
\label{subsec:derivation}

We are interested in the representation of the modular scrambling modes on bulk operators, within the code subspace. This amounts to understanding the contributions to the correlation function
\myeq{\frac{e^{-2\pi \Lambda}}{2\pi}\langle \phi_1\phi_2 \dots [e^{-i\Hm \Lambda}\delta \Hm e^{i\Hm \Lambda}, \phi(x^\alpha, y^i)] \dots \phi_n\rangle \label{bulkGpm}}
with $n\sim O(1)$ that survive in the large $\Lambda$ limit. It is convenient to divide our analysis into three cases depending on the bulk location of the operator $\phi(x^\alpha, y^i)$: (a) Near the RT surface $(x^\alpha K_{ij|\alpha}\ll e^{-2\pi \Lambda})$, (b) far from the RT surface $(x^\alpha K_{ij|\alpha}\gg e^{-2\pi \Lambda})$, (c) the intermediate region $(x^\alpha K_{ij|\alpha}\sim e^{-2\pi \Lambda})$. It is worth noting that in the limit we are interested in $1\ll 2\pi \Lambda \ll \log N$ the region near the RT surface, although exponentially small, is parametrically larger than Planck size and therefore effective field theory applies.

\paragraph{Near the RT surface} By virtue of the JLMS relation \cite{}, the modular Hamiltonian of a CFT subregion is holographically mapped to the bulk QFT modular Hamiltonian of the corresponding entanglement wedge, as long as we are interested in code subspace dynamics. Since the bulk theory is a weakly coupled QFT, $\Hm^{\text{bulk}}$ admits an expansion of the form:
\myal{\Hm^{\text{bulk}}&= \int dx_1 dx_2\, K_{ab}^{(0)}(x_1,x_2) \phi_a(x_1) \phi_b(x_2) \nonumber\\
&+ \frac{1}{N}\int dx_1 dx_2 dx_3 \,K_{abc}^{(1)}(x_1,x_2,x_3) \phi_a(x_1) \phi_b(x_2)\phi_c(x_3) +\dots \label{generalbulkHm}}
where we used $\phi_a$ to denote all the fundamental fields in the theory. While the detailed form of (\ref{generalbulkHm}) depends on the state, its action on localized operators $\phi(x^\alpha, y^i)$ very close to the RT surface $(x^\alpha K_{ij|\alpha}\ll 1)$ is equivalent to a geometric boost on the orthogonal 2D plane, up to $O(x^\alpha K_{ij|\alpha})$ corrections \cite{Faulkner:2017vdd}:
\myeq{\langle \phi_1\phi_2 \dots [\Hm, \phi(x^\alpha, y^i)] \dots \phi_n\rangle =\zeta^\mu_{\text{boost}}\partial_\mu \langle \phi_1\phi_2 \dots  \phi(x^\alpha, y^i) \dots \phi_n\rangle +O(x^\alpha K_{ij|\alpha})\label{Hmboost}}
where $\zeta_{\text{boost}}$ is an approximate local isometry of the bulk metric near the RT surface:
\myeq{\mathcal{L}_{\zeta_{\text{boost}}}g_{\mu\nu}= O(x^2, x^\alpha K_{ij|\alpha})}
with $x^\alpha$ the normal coordinates to the RT surface, located at $x^\alpha=0$. In normal Riemann coordinates about $x^\alpha=0$ (\ref{normalgauge}) the modular boost vector field simply reads:
\myeq{\zeta^\mu_{\text{boost}}\partial_\mu =x^+\partial_+ - x^-\partial_- +O(x^2)}
It follows that $\delta \Hm$ for a shape deformation of the boundary region acts on operators in this neighborhood as the vector flow:
\myeq{\langle \phi_1\phi_2 \dots [\delta \Hm, \phi(x^\alpha, y^i)] \dots \phi_n\rangle =\delta \zeta_{\text{boost}}^\mu(x^\alpha, y^i) \partial_\mu \langle \phi_1\phi_2 \dots  \phi(x^\alpha, y^i) \dots \phi_n\rangle +O(x^\alpha K_{ij|\alpha}) \label{deltaHmbulk}}
The difference of the modular boosts for the two RT surfaces $\delta\zeta_{\text{boost}}^\mu$ was analyzed in \cite{Czech:2019vih}. For clarity, we present an alternative discussion in our Appendix. It is generated by a vector flow $\zeta^\mu$
\myeq{\delta \zeta_{\text{boost}}^\mu= [\zeta^\mu, \zeta_{\text{boost}}^\mu] \label{vectorflow}}
where $[\cdot\,,\cdot]$ is the vector Lie bracket, which can be shown to satisfy (\ref{paralleltransport}) and thus it is a local isometry of the metric at the RT surface, up to $O(x^\alpha K_{ij|\alpha})$ corrections. $\zeta^\mu$ can be decomposed in terms of vector fields with definite weights $\pm 2\pi, 0$ under modular boosts
\myeq{\zeta^\mu= \zeta^\mu_{(+)}+\zeta^\mu_{(-)} +\zeta^\mu_{(0)} \label{decomp}}
The last term in (\ref{decomp}) drops out of the commutator in (\ref{vectorflow}) and (\ref{deltaHmbulk}) becomes:
\myeq{\langle \phi_1\phi_2 \dots [\delta \Hm, \phi(x^\alpha, y^i)] \dots \phi_n\rangle =2\pi \left(\zeta^\mu_{(+)}-\zeta^\mu_{(-)}\right) \partial_\mu \langle \phi_1\phi_2 \dots  \phi(x^\alpha, y^i) \dots \phi_n\rangle +O(x^\alpha K_{ij|\alpha}) \label{deltaHmbulk2}}

When the bulk operator $ \phi(x^\alpha, y^i)$ is located in the region $x^\alpha K_{ij|\alpha}\ll e^{-2\pi \Lambda}$ around the RT surface, the geometric approximation to the bulk modular flow can be trusted for modular times of order $\Lambda$. The action of $G_\pm$ on $\phi$ is obtained simply by using (\ref{deltaHmbulk2}) and the finite version of (\ref{Hmboost}) in (\ref{bulkGpm}):
\myeq{\langle \phi_1\phi_2 \dots [G_\pm, \phi(x^\alpha, y^i)] \dots \phi_n\rangle = \zeta^\mu_{(\pm)} \partial_\mu \langle \phi_1\phi_2 \dots  \phi(x^\alpha, y^i) \dots \phi_n\rangle +O(e^{-2\pi \Lambda}) \label{geometricscramblingmode}}
with the vector fields $\zeta_{(\pm)}$ given in (\ref{scramblingmodeapp}).

\paragraph{Far from the RT surface} For operators away from the edge of the wedge $(x^\alpha K_{ij|\alpha})\gg e^{-2\pi \Lambda})$ modular evolution for $s\sim O(\Lambda)$ will crucially depend on the full, non-local expression (\ref{generalbulkHm}) since the naive application of a Rindler boost moves them to locations with $x^\alpha K_{ij|\alpha}\gg 1$ where the geometric approximation becomes invalid. In computing (\ref{bulkGpm}) we are, thus, required to use the exact modular scrambling modes, which formally are
\myal{G_\pm &=\pm\frac{e^{-2\pi\Lambda}}{2\pi}\Big( \int dx_1 dx_2\, \delta K_{ab}^{(0)}(x_1,x_2) \phi_a(x_1,\Lambda) \phi_b(x_2,\Lambda) \nonumber\\
&+ \frac{1}{N}\int dx_1 dx_2 dx_3\, \delta K_{abc}^{(1)}(x_1,x_2,x_3) \phi_a(x_1,\Lambda) \phi_b(x_2,\Lambda)\phi_c(x_3,\Lambda) +\dots\Big) \label{exactscramblingmodes}}
where we used $\phi(x,\Lambda)= e^{i\Hm \Lambda}\phi(x) e^{-i\Hm \Lambda}$ to avoid clutter. It is generally difficult to say what the commutator $[\phi(x,\Lambda),\phi(x)]$\footnote{inside code subspace correlation functions} is, when $\Lambda\gg 1$, but it is easy to see what \emph{it is not}: exponentially growing with $\Lambda$. Modular flow of a generic state will smear $\phi(x)$ over the interior of the entanglement wedge while preserving its boundedness and, thus, its commutator with all localized bounded operators will typically decay with $\Lambda$ ---a suppression that gets further enhanced by the decaying exponential in the definition of $G_\pm$ (\ref{exactscramblingmodes}). In other words, expressions like (\ref{exactscramblingmodes}) can only lead to exponentially growing contributions under modular flow when the action of the modular operator is geometric. For general states, we therefore expect that
\myal{\langle \phi_1\phi_2 \dots &[G_\pm, \phi(x)] \dots \phi_n\rangle \lesssim O(e^{-2 \kappa \Lambda}) \nonumber\\
\text{for: }& x^\alpha K_{ij|\alpha}\gg e^{-2\pi \Lambda},\nonumber\\
& n\sim O(1)\nonumber\\
& \kappa>0}
An obvious exception to this argument is the case of entanglement wedges for ball shaped boundary subregions in pure AdS. Both the modular Hamiltonian and its shape deformation are simply generators of spacetime's isometries and the modular scrambling modes act geometrically in the entire wedge. This, however, does not contradict our discussion, since it is simply the statement that (\ref{scramblingmodeapp}) is valid in the entire bulk subregion and not just in the RT surface neighborhood.

\paragraph{Intermediate region} The above analysis leaves out a small intermediate region $x^\alpha K_{ij|\alpha}\sim e^{-2\pi \Lambda}$, where the action of $G_\pm$ on bulk operators is not discussed. The difficulty of capturing this edge effect stems from the fact that modular flow remains geometric for almost the entire $s\sim O(\Lambda)$ evolution ---thus generating the enhancement we saw in (\ref{geometricscramblingmode})--- but the non-local corrections become important for an O(1) amount of modular time. $[G_\pm,\phi]$ will therefore contain, besides the geometric piece (\ref{geometricscramblingmode}), non-local contributions which are suppressed by $e^{-O(1)}$ ---an exponent that does not scale with $\Lambda$. As a result we do not have control over $G_\pm$ in this region.

\subsection{Poincare from Lyapunov}
\label{subsec:poincare}

Our discussion reveals that, up to an edge effect, the holographic modular scrambling modes generate a geometric transformation, localized near RT surface. The approximate symmetry (\ref{Gsymmetry}) of the boundary state is mapped to the local symmetry (\ref{localapproxsym}) of the bulk spacetime. 

Most interestingly, the chaos commutator 
\myeq{[\Hm, G_\pm]\approx \pm 2\pi i G_\pm}
which on the boundary is a consequence of the maximal modular ``Lyapunov'' exponent, in the dual gravity picture follows directly from the local Poincare commutator near the RT surface
\myeq{[B, P^\pm]=\pm i P^\pm}
where $B, P^\pm$ stand for the generators of boosts and null translations, respectively. This observation is aligned with the recent discussion of \cite{Lin:2019qwu} where effective symmetry generators of AdS$_2$ were constructed by exploiting the chaotic properties of its SYK dual. It also rhymes with the constructions of \cite{Blake:2017ris} where the maximally chaotic decay of OTOCs was explained, in an effective theory context, as a consequence of a shift symmetry in the effective ``quantum hydrodynamic'' action. 

Similarly, in our work, we see saturation of modular chaos stemming, holographically, from the local Poincare algebra about an entanglement horizon. It is tempting to reverse this logic and ask: Is maximal modular chaos the quantum mechanical principle underlying the emergence of the local translation generators of the dual spacetime?

\subsection{Curvature from Modular Scrambling Algebra}
\label{subsec:commutator}

The geometric flow generated by $G_\pm$ is what was dubbed \emph{modular Berry transport} in \cite{Czech:2019vih, Czech:2018kvg,Czech:2017zfq}. The property that $\zeta_{(\pm)}$ preserve the normal frame of the extremal surface results in a non-trivial algebra for $G_+,G_-$. We may utilize the modular scrambling modes associated to a continuous family of RT surfaces to transport a vector degree of freedom in the bulk. Upon closing this path in the space of RT surfaces, we obtain a non-trivial holonomy transformation which is controlled by the geometric component of the modular Berry curvature \cite{Czech:2019vih}. The scrambling mode commutator underlies this curvature. 

Using the explicit form for the flows $\zeta_{(\pm)}$, the scrambling commutator can be explicitly computed (see Appendix). In normal Riemann gauge about the RT surface (\ref{normalgauge}) we find:
\myal{ [G_+,G_-] &= J^i(y) \partial_i  +\delta x^+ \delta x^- R_{+-|}\,^{\mu}\,_{\alpha}\,x^\alpha \partial_\mu \nonumber\\
&+\left( \frac{1}{2}\epsilon_{-+}\nabla^i \delta x^- \nabla_i \delta x^+-2J^i(y)a_i(y) \right)(x^+\partial_+ -x^-\partial_-)\nonumber\\
\text{where: }&J^i= \frac{1}{2}\,\left(\delta x^- \nabla^i \delta x^+ - \delta x^+ \nabla^i \delta x^-\right)
\label{Gcommutator}}
with $\nabla_i \delta x^\alpha =\partial_i \delta x^\alpha  +\Gamma^\alpha_{i\beta}\delta x^\beta= \partial_i \delta x^\alpha  +a_i \eta^{\alpha \gamma}\epsilon_{\beta\gamma} \delta x^\beta$ and $\delta x^\pm(y)$ the lightlike components of the vector field describing the RT surface deformation. The latter is a solution to the differential equation
\myeq{\eta_{\alpha\beta} \nabla^i\nabla_i \delta x^\beta = R^i_{(\alpha| \beta)i} \delta x^\beta -K_{\alpha|ij}K_{\beta}^{ij} \delta x^{\beta}}
which follows from the requirement that the deformed surface must also be extremal \cite{Lewkowycz:2018sgn}. Eq. (\ref{Gcommutator}) is not an operator equality but, as in the previous section, the statement that correlation functions of $[[G_+,G_-],\phi(x)]$ for $x\ll e^{-2\pi \Lambda} \ell$ are obtained by acting with the differential operator in the r.h.s of (\ref{Gcommutator}) on the correlation functions of $\phi(x)$. Moreover, since we have neglected corrections of the order $(x^\alpha K_{ij|\alpha})$ throughout the bulk discussion, the commutator (\ref{Gcommutator}) can be consistently considered non-vanishing only when there is a hierarchy between the extrinsic curvature of the RT surface and the bulk curvature:
\myeq{R \gg K^2}

The geometric representation of the bulk $G_\pm$ establishes a direct link between the $[G_+,G_-]$ and the bulk curvature. This clarifies the proposal of \cite{Czech:2019vih} and highlights an aspect of entanglement, encoded in the modular scrambling mode algebra, that reflects local gravitational data.

\section{Outlook}
\label{sec:conclusion}
The puzzle of bulk reconstruction is a question of principle. Bulk quantities can often be re-expressed in the boundary language but satisfactory developments ought to do more: They must point at those quantum principles that underly the effective gravitational organization of the CFT data and reflect the key features of the spacetime description. In this paper, we offer an observation of this kind, relating the emergent algebra of local shifts about a Ryu-Takayanagi surface to the saturation of a Quantum Field Theory bound on modular flow which we conjectured and proved in certain cases. An interesting direction for future work is to try to improve on our arguments in Section \ref{subsubsec:state} and derive the bound for general state deformations. The ideas of \cite{araki2} may prove helpful in this task.

\paragraph{Relation to operator size} Interesting recent works \cite{Susskind:2018tei, Qi:2018bje} have suggested the identification of the radial momentum of a moving particle in AdS$_2$ with a notion of ``size'' of the corresponding operator in the boundary quantum system. The key observation in favor of this proposal crucially relies on the maximally chaotic nature of the boundary dual: The size of the boundary excitation grows exponentially in time, closely resembling the exponential blueshifting of a bulk particle as it freely falls towards the AdS$_2$ black hole. Based on this proposal, effective generators of bulk translations can be constructed \cite{Lin:2019qwu}. 

It is interesting to explore the connection of these ideas to the results presented here. The notion of modular chaos developed in this work may serve as a starting point for defining operator size in higher dimensional theories. Conversely, thinking in terms of the modular Hamiltonians for the degrees of freedom carried along an observer's worldline, instead of the subalgebras of boundary subregions, may link modular chaos to the Poincare algebra in a local bulk neighborhood.

\paragraph{Effective theory for modular chaos and gravity} From the perspective of the approximately flat region in the vicinity of a Ryu-Takayanagi surface, our modular scrambling modes look like light-ray operators of the bulk stress tensor. These operators are closely related to gravitational physics. In the '90s, 't Hooft \cite{tHooft:1996rdg} and Erik and Herman Verlinde \cite{Verlinde:1991iu} formulated classical gravitational scattering of highly boosted particles in the vicinity of a black as an effective theory for such light-ray operators. This hints at the intriguing possibility of capturing gravitational scattering around an RT surface via an effective theory for modular chaos, perhaps closely related to the discussions of \cite{Blake:2017ris, Haehl:2018izb, Haehl:2019eae} ---a direction we plan to explore in the future. The properties of these light-ray operators were also studied more recently in \cite{Kologlu:2019bco, Kologlu:2019mfz, Belin:2019mnx} where the commutativity of parallel shockwaves was linked to Einstein gravity. The connection of these works to the results presented here is a direction worth investigating further.

\acknowledgments{LL is grateful to Daniel Jafferis for many enlightening conversations and related work.  We give special thanks to Nima Lashkari for his useful feedback and suggestions. Furthermore, we want to thank Nele Callebaut, Bartek Czech, Netta Engelhardt, Tom Faulkner, Ben Freivogel, 
Daniel Harlow, Gilad Lifschytz, Hong Liu, Moshe Rozali, Sasha Zhiboedov, Claire Zukowski for useful conversations. We both benefitted from the  ``String Theory workshop'' at the University of Amsterdam and the ``10th Crete Regional String meeting''. LL greatly appreciates the hospitality of Technion and Haifa Universities. LL is supported by the Pappalardo Fellowship. JdB is supported by the European Research Council under the European
Union's Seventh Framework Programme (FP7/2007-2013), ERC Grant agreement
ADG 834878.}

\appendix
\section{Bulk modular boosts and scrambling modes}
The normal Riemann gauge about a codimension-2 extremal surface takes the form:
\myal{ g_{\alpha\beta} &= \eta_{\alpha \beta} -\frac{1}{3}R_{\alpha\gamma|\beta\delta} x^\gamma x^\delta +O(x^3)\\
g_{\alpha i} &= a_i(y) \epsilon_{\gamma\alpha}x^\gamma +\frac{2}{3} R_{i\beta|\gamma\alpha} x^\gamma x^\beta +O(x^3)\\
g_{ij}&= \gamma_{ij} +2x^{\alpha}K_{ij|\alpha} + \left( -4\eta_{\alpha\beta} a_i a_j +R_{i(\alpha|\beta) j} +K_{\alpha|il} K_{\beta|jk} \gamma^{lk} \right)x^{\alpha}x^\beta +O(x^3) \label{normalgauge}
}
The geometric flow generated by the modular Hamiltonian in the region $x^\alpha K_{ij|\alpha}\ll 1$ is a boost in the orthogonal 2D plane to the surface:
\myeq{\zeta_{\text{boost}}^\mu\partial_\mu= x^+\partial_+-x^-\partial_- +O(x^2)}
Since, quantum mechanically $\Hm|\psi\rangle=0$ this geometric boost must be an isometry of the metric in this region. It can be checked explicitly that, indeed,
\myeq{\mathcal{L}_{\zeta_{\text{boost}}}g_{\mu\nu} = O(x^2) +O(x^\alpha K_{ij|\alpha}) \label{killingapp}}

Consider now an infinitesimally separated minimal surface, located at: $x^\alpha= \delta x^\alpha(y)$. The modular boost of the new subregion $\zeta'_{\text{boost}}$ is again defined as the approximate killing vector (\ref{killingapp}) whose leading order behavior is a boost in the new surface's normal plane. The two vector fields are related by a vector flow $\zeta$ which at leading order acts the appropriate transverse shift of the surface by $\delta x^\alpha(y)$ and solves the equation
\myeq{\delta \zeta_{\text{boost}}=[\zeta, \zeta_{\text{boost}}] \label{vecflowapp}}
Since both $\zeta_{\text{boost}}$ and $\zeta_{\text{boost}}'$ act as normal boosts near the two surfaces, $\zeta$ can alternatively be obtained as the map between the two different orthonormal coordinate frames:
\myal{\mathcal{L}_{\zeta}g_{\alpha\mu}(x^\alpha=0,y^i) &= 0 \label{killingscramblingapp}}
which can be viewed as an alternative definition of the vector field $\zeta$. Combined with the fact that $\mathcal{L}_{\zeta}g_{ij}(x^\alpha=0,y^i) = O(\zeta^\alpha K_{ij|\alpha})$, this vector field generates an approximate isometry in the vicinity of the surface, which is consistent with the invariance of the dual state (\ref{Gsymmetry}). The general solution to (\ref{killingscramblingapp}) in the gauge (\ref{normalgauge}) reads, up to $O(x^2, x^\alpha K_{ij|\alpha})$ corrections:
\myal{\zeta^{\alpha}&= \delta x^{\alpha}+O(x^2)\nonumber\\
\zeta^i&=-\left(\gamma^{ij} \eta_{\alpha\beta}\nabla_j \delta x^\alpha\right)x^\beta +O(x^2)\nonumber  \\
\text{where: } &\nabla_i \delta x^\alpha =\partial_i \delta x^\alpha  +\Gamma^\alpha_{i\beta}\delta x^\beta= \partial_i \delta x^\alpha  +a_i \eta^{\alpha \gamma}\epsilon_{\sigma\gamma} \delta x^\sigma
\label{vectorfields}}
    The vanishing of $\partial_\beta \zeta^\alpha$ is a consequence of the fact that $\Gamma^\mu_{\alpha\beta}=0$ in the normal Riemann gauge (\ref{normalgauge}). In more general gauge this term is non zero and it depends on the Christoffel symbols. The expression in general gauge can of course be obtained by a diffeomorphism of the solution (\ref{vectorfields}). The vector field (\ref{vectorfields}) is not unique, as discussed at length in \cite{Czech:2019vih} but a ``gauge''-equivalence class of solutions may be generated by simply adding to $\zeta$ contributions that commute with $\zeta_{\text{boost}}$. These contributions, however, drop out from (\ref{vecflowapp}) and do not affect the modular scrambling modes.

The modular scrambling modes are defined as the contributions to $\delta \Hm$ that saturate the modular chaos bound. Close to the RT surface, modular flow is simply a normal boost (\ref{killingapp}) and (\ref{vectorfields}) decomposes, straightforwardly, into a sum of two vector fields with boost weight $+1$ and $-1$ respectively:
\myal{ \zeta_{(+)}^\mu\partial_\mu = \delta x^+ \partial_+ - x^-\gamma^{ij} \nabla_j \delta x^+ \partial_i +O(x^2)\nonumber\\
 \zeta_{(-)}^\mu\partial_\mu = \delta x^- \partial_- - x^+\gamma^{ij} \nabla_j \delta x^- \partial_i +O(x^2) \label{scramblingmodeapp}}
These are, therefore, the contributions that dominate at large positive and large negative modular times respectively.

\paragraph{Scrambling mode commutator}
The scrambling mode commutator $[G_+, G_-]$ can be represented on bulk operators close to the RT surface as the vector flow generated by the Lie bracket of $\zeta_{(\pm)}$:
\myeq{[G_+, G_-] \rightarrow [\zeta_{(+)},\zeta_{(-)}]^\mu \partial_\mu  \label{vectorcommutator}}
In performing this computation, one needs to pay attention to a subtlety. The modular scrambling modes expressions (\ref{vectorfields}) are valid in the normal Riemann gauge where $\Gamma^\mu_{\alpha\beta}=0$. However, acting with one of these vector fields will move the surface and change the gauge, as can be seen by (\ref{normalgauge}). This in turn affects the expression for the modular scrambling mode we act with next. A convenient way to deal with this complication is to adjust the $O(x^2)$ contributions to the approximate  Killing vectors (\ref{vectorfields}) so that they preserve the local gauge and, thus allowing us to use (\ref{scramblingmodeapp}) for the second application of the modular scrambling mode in (\ref{vectorcommutator}). These gauge-preserving extended vector fields read:
\myal{\zeta^{\alpha}&= \delta x^{\alpha}+\left(\frac{1}{3} \delta x^{\lambda} \eta^{\alpha\delta}R_{\beta(\lambda|\delta)\gamma} +\frac{1}{2} a_i \nabla^i\delta x^\lambda(\epsilon_{\beta\sigma}\,\eta^{\sigma\alpha}\eta_{\gamma\lambda}+\epsilon_{\gamma\sigma}\,\eta^{\sigma\alpha} \eta_{\beta\lambda}) \right) x^{\beta}x^\gamma\nonumber\\
\zeta^i&=-\left(\gamma^{ij} \eta_{\alpha\beta}\nabla_j \delta x^\alpha\right)x^\beta + \frac{1}{3} \delta x^\beta R^i\,_{\alpha|\gamma\beta} x^\alpha x^\gamma \\
\text{where: } &\nabla_i \delta x^\alpha =\partial_i \delta x^\alpha  +\Gamma^\alpha_{i\beta}\delta x^\beta= \partial_i \delta x^\alpha  +a_i \eta^{\alpha \gamma}\epsilon_{\sigma\gamma} \delta x^\sigma
\label{extendedvectorfields}}
Their commutator (\ref{Gcommutator}) follows by a straightforward computation.


%
%
%
%

\end{document}